# Emergence String and Mass Formulas of Hadrons

Yi-Fang Chang

Department of Physics, Yunnan University, Kunming 650091, China

(e-mail: yifangchang1030@hotmail.com)

**Abstract:** Assume that hadrons are formed from the emergence string. Usual string should possess two moving states: oscillation and rotation, so we propose corresponding potential and the equation of the emergence string, whose energy spectrum is namely the GMO mass formula and its modified accurate mass formula $M = M_0 + AS + B[I(I+1) - S^2/2]$. These are some relations between the string and observable experimental data.

### 1. Introduction

In 1969, Nambu, Nielson and Susskind proposed a string model, whose basis is that the excited spectrum of the dual model of hadrons is mathematically analogous to a string, so all particles are various excited states of the basic string. Such the free string is used to build the operator formalism of dual models, and the string theory becomes a new picture of various types of hadrons, as different modes of vibration of a string [1-3]. The simplest form in superstring theories is in 10 spacetime dimensions [4,5]. Green and Schwarz proposed the D=10 superstring theory [6].

Englert, et al., clarified the mechanism of the emergence of the internal gauge symmetry as well as the space-time fermions and supersymmetry [7], and explained the states, operators and interactions of superstring emerge from this theory [8]. Szczepaniak, et al., investigate a relativistic many-body approach to hadron structure based on the Coulomb gauge QCD Hamiltonian, in which dynamical chiral symmetry breaking naturally emerges, and both quarks and gluons acquire constituent masses when standard many-body techniques are employed [9]. Teramond discussed the fundamental chiral nature of the observed quarks and leptons and the emergence of the gauge group of the standard model [10]. Wen demonstrated that a condensed state of nets of strings, and the gauge bosons and fermions can emerge from a local bosonic model [11]. Carlson, et al., studied a relativistic quantum model of particles with internal structure, which is suggested by the string model [12]. Frampton discussed the two string approaches to hadron structure, which are related with the vector mass and the phenomenological linearity of the Regge trajectory [13]. Gupta and Rosenzweig studied the decay of hadrons based on a semiclassical string model, and found that the width to mass ratio $\Gamma/m$ is an increasing function of m, and the decay probability of hadrons on the leading Regge trajectories is computed taking the effect of the string rotation into account [14]. Kobayashi obtained an instructive pattern of quark mass matrices [15]:

$$M_{u,d} = c_{u,d} \begin{pmatrix} \varepsilon_{11} & \varepsilon_{u,d}^3 & \varepsilon_{13} \\ \varepsilon_{u,d}^3 & \varepsilon_{u,d}^2 & \varepsilon_{u,d}^2 \\ \varepsilon_{13} & \varepsilon_{u,d}^2 & 1 \end{pmatrix}. \qquad (1)$$

Feigenbaum, et al., proposed a string picture of hadrons, and led to vector meson masses and so on, and is endowed with an extra discrete space dimension involving a minimal two-point lattice with a spacing of order $10^{-14}$ cm in one of the two *T*-duality [16]. Bigazzi, et al., computed the one-loop string corrections to the Wilson loop, glueball Regge trajectory and stringy hadron masses in a Yang-Mills theory from semiclassical strings, and the one-loop corrections to the linear glueball Regge trajectories [17].

Recently, Konopka, et al., presented a model of emergent locality, which can give rise to an emergent U(1) gauge theory by the string-net condensation mechanism [18]. L?wen, et al., studied the models with a doubly suppressed gravitino mass, which can emerge from a condensate as large as the grand unified scale, and analyzed the properties of these models and discuss applications for particle physics and cosmology [19]. Cvetič and Langacker presented a stringy mechanism to generate Dirac neutrino masses by D-instantons in string theory [20]. Roy, et al., shown that how string resonances, whose formation is an exciting possibility of discover new physics beyond SM, may be detected at the LHC in the $pp \to \gamma + $ jet channel [21].

The emergence string even may become macroscopic object. Perry and Teo investigate that in the semiclassical limit, the two-dimensional space-time which emerges from string theory is a black hole similar to the Schwarzschild solution, and calculated the mass and temperature associated with this space-time [22]. Brustein and de Alwis proposed the universe emerged from the string era in a thermally excited state, and discussed the landscape of string theory and the wave function of the universe [23]. Battefeld and Shuhmaher searched the predictions of dynamically emerging brane inflation models [24]. Niz and Turok studied the classical evolution of fundamental strings across a big crunch/big bang singularity [25]. But, usual string theory does not consider and calculate many concrete questions in particles.

On the other hand, it is very successful that the SU(3) symmetry and its broken are applied to the classification of the ground hadrons, which made of u, d and s quarks, and to the mass spectrum and so on. It is also well known that those hadrons agree with the GMO mass formula

$$M = M_0 + AS + B[I(I+1) - \frac{S^2}{4}]. \qquad (2)$$

For the SU(8) supermultiplets there is a simple mass formula [26]:

$$M = M_0 + \alpha Y + \beta C + \gamma[\frac{Y^2}{4} - I(I+1)] + \delta J(J+1). \qquad (3)$$

What is the relation between string and particles and their properties (mass, charge, spin and so on)?

**2. Mass formulas derived from the emergence string**

Usual string should possess two moving states: oscillation and rotation. Iwasaki and Takagi discussed classical theory and quantum theory of the string motion, and presented a relativistic string model for hadrons, in which the vibrational and rotational motions of the string are solved by using the WKB approximation. Here all the hadrons have universal stringlike structure, they calculated mass spectra and decay widths of $\rho$ and K* resonances [27]. Rey and Sugimoto investigated the real-time tachyon dynamics of an unstable D-brane carrying fundamental string charge, and the rolling of the modulated tachyon with gauge flux and an emergent fundamental string [28]. Shifman and Yung considered the main feature of the non-Abelian strings associated with the possibility of rotations of their color fluxes inside a global SU($N$) group [29]. Levin and Wen gave an example of a purely bosonic rotor model on the 3D cubic lattice, whose low energy excitations behave like massless U(1) gauge bosons and massless Dirac fermions, in which both photons and electrons. It illustrates a general mechanism for the emergence of gauge bosons and fermions known as string-net condensation, and string-net condensed models can have excitations that behave like gluons, quarks and other particles in the standard model, so photons, electrons and other elementary particles may have a unified origin: string-net condensation in this vacuum [30].

Based on the general oscillation and rotation of string we propose that a corresponding equation of the emergence string may be

$$\frac{d^2 S}{dr^2} + [-\frac{K(K+1)}{r^2} + 2m(E'-U)]S = 0. \qquad (4)$$

Here S may be a wave function or other function. Moreover, this equation may regard one of a scalar field in the gauge theory derived from emergence string [18]. If U is a potential of the anharmonic oscillator [31,32]:

$$U(r) = \frac{1}{2}m\omega^2 r^2 + \alpha r^3. \qquad (5)$$

For example, Bolokhov, et al., described the heterotic string solutions in the M model, which obtained in U($N$) gauge theories for N=2 supersymmetric QCD deformed by superpotential terms $\mu A^2$ breaking [33]. Polyakov constructed vertex operators for massless higher spin fields in Ramond-Neveu-Schwarz superstring theory, and derived the gauge-invariant cubic interaction terms for the higher spins, in which the $\psi$ part is $<\psi^{a_3}(z)\partial\psi^{b_3}\psi^{b_4}(w)\psi^{c_3}(u)>$ [34].

Such a corresponding energy level for Eq.(4), in second approximation, is:

$$E_{k,n} = E_{0,0} + \hbar\omega(n+\frac{1}{2}) + \frac{\hbar^2}{2mr_0^2}K(K+1) - \frac{\hbar^2\alpha^2}{8m^3\omega^4}(30n^2+30n+11). \quad (6)$$

It is namely

$$E_{k,n} = E'_{0,0} + (\hbar\omega - \frac{15\hbar^2\alpha^2}{4m^3\omega^4})n + \frac{\hbar^2}{2mr_0^2}K(K+1) - \frac{15\hbar^2\alpha^2}{4m^3\omega^4}n^2. \quad (7)$$

If oscillation and rotation of the emergence string correspond respectively to quantum numbers S and I, this energy level will correspond completely to the mass formula of hadrons:

$$M = M_0 + AS + BI(I+1) + CS^2. \quad (8)$$

When C=-B/4, it is the GMO mass formula (2). When C=-B/2, it is a modified accurate mass formula [35,36]

$$M = M_0 + AS + B[I(I+1) - \frac{S^2}{2}]. \quad (9)$$

For the $J^P = 1^+/2$ baryon octet, let $M_0 = 908, A = -228,$ and B=40MeV, so

$$m(N) = 938, m(\Lambda) = 1116, m(\Sigma) = 1196, m(\Xi) = 1314 \text{ MeV}. \quad (10)$$

Since S<0 correspond to n>0, so $\hbar\omega = 208$MeV, $\omega = 3.160 \times 10^{23} s^{-1}$, and the rotational inertia of emergence string is $J = mr_0^2 = \hbar^2/80 = 5.4155 \times 10^{-45} MeVs^2$, assume that mass is one of u quark $m_u \approx 3 MeV/c^2 = 3.333 \times 10^{-21} MeVcm^{-2}s^2$, so $r_o = \sqrt{J/m} = 1.275 \times 10^{-12} cm$ is a rotation scale of baryons formed from emergence string. The scale of string is very small $10^{-33}$ cm.

For the $J^{pc} = 0^{-+}$ meson octet, let A=0, $M_0$ =549.4MeV and $B = -207.22$ MeV, so

$$m(\pi^0) = 134.96, m(K^0) = 497.6, m(\eta) = 549.4 \text{ MeV}. \quad (11)$$

The neutral mesons agree completely within the range of error, and M=m. Since A=0 correspond to $\hbar\omega = B/2 = -103.61$ MeV, $\omega = 1.574 \times 10^{23} s^{-1}$ is about a half of baryon, and the rotational inertia of emergence string is $J = mr_0^2 = \hbar^2/414.44 = 1.0454 \times 10^{-45} MeVs^2$, assume that mass is also one of u quark, so $r_o = 3.136 \times 10^{-13} cm$ is a rotation scale of mesons formed from emergence string, which is about a quarter of baryon. Such masses of mesons are smaller than baryons, which may be attributed to smaller oscillation and rotation of emergence string.

The mass relations

$$2[m(N) + m(\Xi)] = 3m(\Lambda) + m(\Sigma), \quad (12)$$

$$4[m(N) + m(\Xi)] = 7m(\Lambda) + m(\Sigma), \quad (13)$$

$$8m(K) = 7m(\eta) + m(\pi). \quad (14)$$

They correspond to Eqs.(2) and (9), respectively.

For the $J^P = 3^+/2$ baryon decuplet I=1+(B+S)/2 always holds, so the formula (9) just derives a more accurate formula

$$M = M_0 + aS + bS^2 = M'_0 + AY + CY^2. \quad (15)$$

Let $M'_0 = 1385.3, A = -150.2$ and $C = -3.3$ MeV, so

$$m(\Delta) = 1231.8, m(\Sigma) = 1385.3, m(\Xi) = 1532.2, m(\Omega) = 1672.5 \text{ MeV}. \quad (16)$$

It can suppose to be a result of the anharmonic oscillator without rotation.

If a potential of the anharmonic oscillator [31,32] is:

$$U(r) = \frac{1}{2}m\omega^2 r^2 + \alpha r^3 + \beta r^4, \quad (17)$$

its energy level and corresponding mass formula will have a term of $n^3$ and $S^3$.

Further, in the standard model quarks are the three generations (u,d) (c,s) and (t,b), whose properties exhibit a better symmetry. According to the symmetry of s-c quarks in the same generation, the heavy flavor hadrons which made of u,d and c quarks should be also classified by SU(3) octet and decuplet. Such some simple mass formulas are obtained [37], from this we can predict some masses of unknown hadrons, for example, $m(\Xi_{cc}) = 3715$ or $3673 MeV$, and m($\Omega_{cc}^+$)=3950.8 or 3908.2MeV, etc.

Moreover, in Eq.(5) if the potential is the Morse-type potential [31,36] or the infinite square well potential in spherical coordinates [38], the same energy level will be also obtained approximately. A general radial wave equation with rotationless is also Eq.(4) [38].

Braaten, et al., studied nonperturbative weak coupling analysis of the Liouville quantum field theory, and the final term in the string equation

$$(\partial^2/\partial\sigma^2 - \partial^2/\partial\tau^2)\varphi = (4m^2/g)e^{2g\varphi}. \qquad (18)$$

It is analogue with the Morse potential [39]. Kikkawa, et al., studied a semiclassical approach to the quark-string model and the hadron spectrum, and provided a relationship between the asymptotic Regge slope and the hadron structure, in which the center of mass of the string is shown to play a special role in classical solutions [40]. Some string theories derived also the Regge trajectory [13,14,17]. For various resonances, the potential of the emergence string may be simplified to a harmonic oscillator, such the corresponding energy level is namely the Regge formula S=AJ+B.

These are some relations between the string and observable experimental data. Generally, the above results all hold for the brane extended from string, since the branes may also have oscillation and rotation.

**References**
1. P.Ramond, Phys.Rev. D3,2415(1971).
2. A.Neveu and J.H.Schwarz, Nucl.Phys. B31,86(1971); Phys.Rev. D4,1109(1971).
3. J.Scherk, Rev.Mod.Phys. 47,123(1975).
4. J.H.Schwarz, Nucl.Phys. B46,61(1972).
5. E.Witten, Nucl.Phys. B445,85(1995).
6. M.B.Green and J.H.Schwarz, Nucl.Phys. B181,502(1981); Phys.Lett. 149B,117(1984).
7. F.Englert, A.Neveu, Phys.Lett. 163B, 349(1985).
8. F.Englert, H.Nicolai, A.Schellekens, Nucl.Phys. B274,315(1986).
9. A.Szczepaniak, E.S.Swanson, C.-R.Ji and S.R.Cotanch, Phys.Rev.Lett. 76,2011(1996).
10. G.F.de Teramond, Phys.Rev. D60,095010(1999).
11. Xiao-Gang Wen, Phys.Rev. D68,065003(2003).
12. C.E.Carlson, L.N.Chang, F.Mansouri and J.F.Willemsen, Phys.Rev. D10,4218(1974).
13. P.H.Frampton, Phys.Rev. D12,538(1975).
14. K.S.Gupta and C.Rosenzweig, Phys.Rev. D50,3368(1994).
15. T.Kobayashi, Phys.Lett. B358,253(1995).
16. J.A.Feigenbaum, P.G.O.Freund and M.Pigli, Phys.Rev. D56,2590(1997).
17. F.Bigazzi, A.L.Cotrone, L.Martucci and L.A.P.Zayas, Phys.Rev. D71,066002(2005).
18. T.Konopka, F.Markopoulou and S.Severini, Phys.Rev. D77,104029(2008).
19. V.L?wen, H.P.Nilles and A.Zanzi, Phys.Rev. D78,046002(2008).
20. M.Cvetič and P.Langacker, Phys.Rev. D78,066012(2008).
21. A.Roy and M.Cavaglia, Phys.Rev. D80,015006(2009).
22. M.J.Perry and E.Teo, Phys.Rev.Lett. 70,2669(1993).
23. R.Brustein and S.P.de Alwis, Phys.Rev. D73,046009(2006).
24. T.Battefeld and N.Shuhmaher, Phys.Rev. D74,123501(2006).
25. G.Niz and N.Turok, Phys.Rev. D75,026001(2007).
26. J.W.Maffat, Phys.Rev. D12,288(1975).
27. M.Iwasaki and F.Takagi, Phys.Rev. D59,094024(1999).
28. S.-J.Rey and S.Sugimoto, Phys.Rev. D68,026003(2003).
29. M. Shifman and A. Yung, Phys. Rev. D 72, 085017 (2005).
30. M.Levin and X.-G.Wen, Phys.Rev. B73,035122(2006).
31. L.D.Landau, E.M.Lifshitz, Quantum Mechanics, Non-Relativistic Theory. Oxford: Pergamon Press. 1977.
32. S.Flugge, Practical Quantum Mechanics. Springer-Velag Berlin Heidelberg. 1999.
33. P.A.Bolokhov, M.Shifman and A.Yung, Phys.Rev. D79,106001(2009).

34. D.Polyakov, Phys.Rev. D82,066005(2010).
35. Yi-Fang Chang, Hadronic J. 7,1118,(1984).
36. Yi-Fang Chang, New Research of Particle Physics and Relativity. Yunnan Science and Technology Press. (1989)p1-183; Phys.Abst., 93(1990)No1371.
37. Yi-Fang Chang, arXiv:0803.0087.
38. E.S.Abers, Quantum Mechanics. Pearson Education, Inc. 2004. p95.
39. E.Braaten, T.Curtright, G.Ghandour and C.B.Thorn, Phys.Rev.Lett. 51,19(1983).
40. K.Kikkawa, T.Kotani, M.Sato and M.Kenmoku, Phys.Rev. D19,1011(1979).